\documentclass[runningheads]{ASMDA}

 \usepackage{mwe}
 \usepackage{eurosym}
 \usepackage{color}
 \usepackage{graphicx} 
 \usepackage{subfig}
 \usepackage{makecell}
\usepackage{cite}

\begin{document}

\title*{Forecasting wind speed financial return}
\toctitle{Forecasting wind speed financial return}
%
\titlerunning{Forecasting wind speed financial return}

\author{
Guglielmo D'Amico\inst{1}
\and
  Filippo Petroni\inst{2}
  \and 
  Flavio Prattico\inst{3}
}
%
\index{D'Amico, G.}
\index{Petroni, F.}
\index{Prattico, F.}

%
\authorrunning{D'Amico et al.}
%
\institute{
Dipartimento di Farmacia, 
Universit\`a `G. D'Annunzio' di Chieti-Pescara,  66013 Chieti, Italy\\
(E-mail: {\tt gdamico@unich.it})
\and
  Dipartimento di Scienze Economiche ed Aziendali,
  Universit\`a degli studi di Cagliari, 09123 Cagliari, Italy\\
  (E-mail: {\tt fpetroni@unica.it})
   \and 
 Dipartimento di Ingegneria Industriale e dell'Informazione e di Economia,\\
 Universit\`a degli studi dell'Aquila, 67100 L'Aquila, Italy\\
  (E-mail: {\tt flavio.prattico@univaq.it})
}

\maketitle             

\begin{abstract}
The prediction of wind speed is very important when dealing with the production of energy through wind turbines. In this paper, we show a new nonparametric model, based on semi-Markov chains, to predict wind speed. Particularly we use an indexed semi-Markov model that has been shown to be able to reproduce accurately the statistical behavior of wind speed. The model is used to forecast, one step ahead, wind speed. In order to check the validity of the model we show, as indicator of goodness, the root mean square error and mean absolute error between real data and predicted ones. We also compare our forecasting results with those of a persistence model. At last, we show an application of the model to predict financial indicators like the Internal Rate of Return, Duration and Convexity.
\end{abstract}

\section{Introduction}
The variations of wind speed, in a certain site, are strictly related to the economic aspects of a wind farm, such as maintenance operations, especially in the off shore farms, pitch angle control on new wind turbines and evaluation of a new site. Many scholars have proposed new models that can allow the prediction of wind speed, minutes, hours or days ahead. Many of these models are based on neural networks \cite{biv}, autoregressive models \cite{pog}, Markov chains \cite{sha}, hybrid models where the previous mentioned models are combined \cite{pou}, \cite{liu} and other models \cite{sal,seg,tar}. Often, these models are either focused on specific time scale forecasting, or synthetic time series generation. 

The approach we propose here is based on indexed semi-Markov chain (ISMC) model that was advanced by the same authors in \cite{wind2} and applied to the generation of synthetic wind speed time series. In \cite{wind2} we showed that our model is able to reproduce correctly the statistical behavior of wind speed. The ISMC model is a nonparametric model because it does not require any assumption on the form of the distribution function of wind speed.
In this work we use the same model to forecast future values of wind speed. We will show that this model performs better than a simple persistence model, by comparing the root mean square errors and the mean absolute errors. The  ISMC model is able to forecast wind speed at different time scale without loosing the goodness of forecasting which is almost independent from the time horizon. 

The economic aspect of the wind turbine energy production is taken into account by evaluating financial indicators of an investment in wind farm (Internal Rate of Return, Duration and Convexity). We also consider two different cases: the first one by using the energy price adding the government incentives and the second without them. 

The paper is organised as follows: first, in Section 2, we describe the database and the used commercial wind turbine. Next, in Section 3, basic notation and definitions on the indexed semi-Markov chain model are provided. Section 4 is the most important part of the paper; it contains the relevant results about wind speed forecasting. Continuing, Section 5 demonstrates the economic application of the model to a real dataset. Finally, Section 6 presents some concluding remarks.  

\section{Database and commercial wind turbine}
The database used in this work is the same used in our precedent work \cite{wind2,wind1,wind4,wind4b,wind5,wind3}, in which it is possible to find a more accurate description and additional information. In this analysis, for the validation of the forecasting model and in order to analyze the behavior at different time scales, we need to resample the original data at different sampling frequencies: namely 30 minutes, 1 hour and 2 hours.

In this work we are going to use the same database also to test our model in the production of energy from a commercial wind turbine which has the blades at a given altitude. Then, we need to consider that the altitude from the ground influences the speed of wind. In particular, it is well known in the literature (see \cite{gas}) that the wind speed has the following dependence from the altitude: 
\begin{equation}\label{height}
v_h = v_{rif} \left( \frac{h}{h_{rif}} \right)^{\alpha} \;\;\;\;\; \alpha= \frac{1}{ln\frac{h}{z_0}}
\end{equation}
where $v_h$ is the wind speed at the height of the wind turbine hub, $v_{rif}$ is the value of the wind speed at the height of the instrument, $h$ and $h_{rif}$ are respectively the height of the wind turbine and of the instrument ($h=50m$ and $h_{rif}=22m$) and $z_0$ is a factor that takes into account the morphology of the area near the wind turbine. This parameter for a region in which there are no buildings or trees varies from 0.01 to 0.001, instead for the offshore application it is equal to 0.0001. For our analysis we consider a mean value for an onshore application, then we choose $z_0 =0.005$.

\begin{figure}
\centering
\includegraphics[height=8cm]{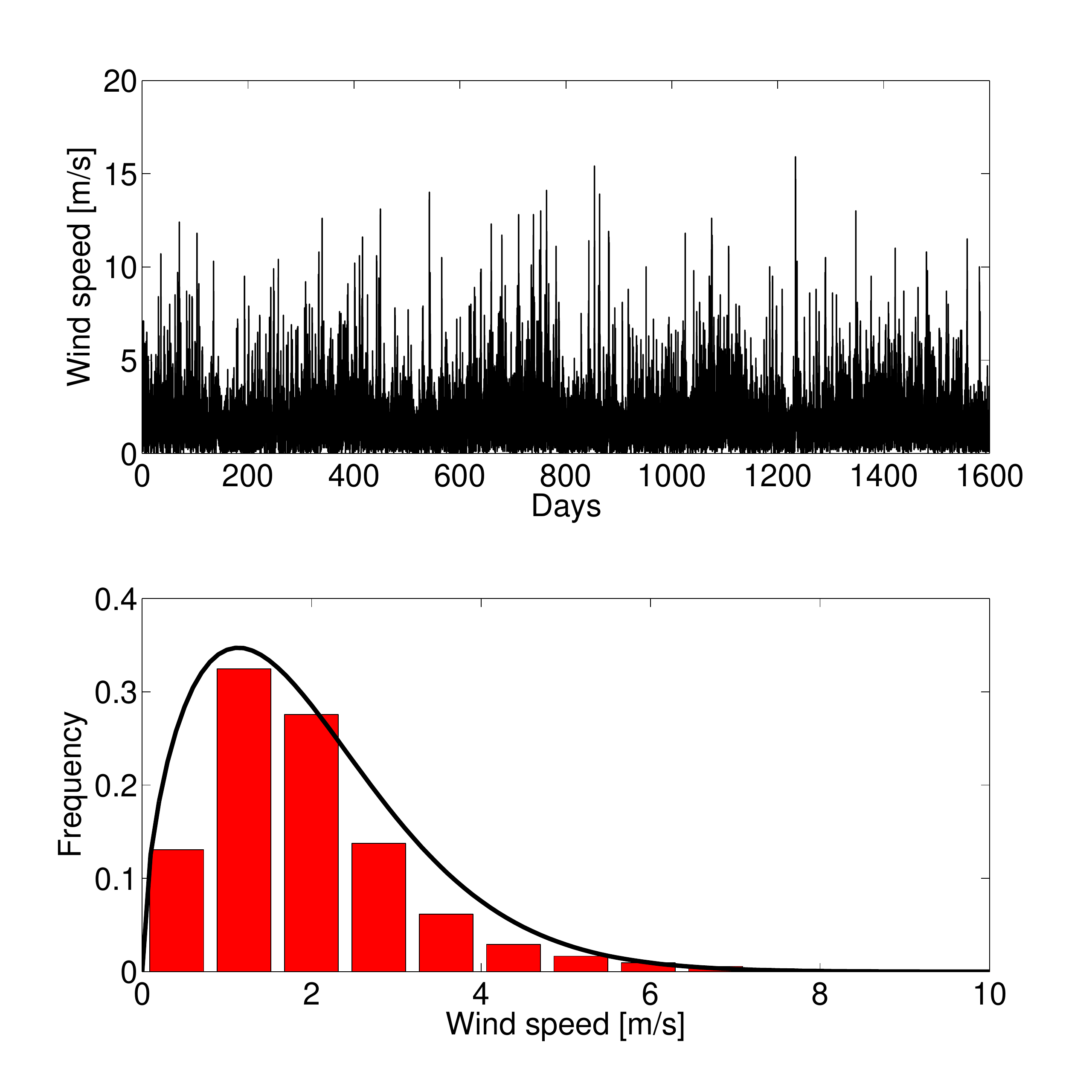}
\caption{Database and its probability density distribution.}\label{fig1}
\end{figure}

\begin{figure}
\centering
\includegraphics[height=6cm]{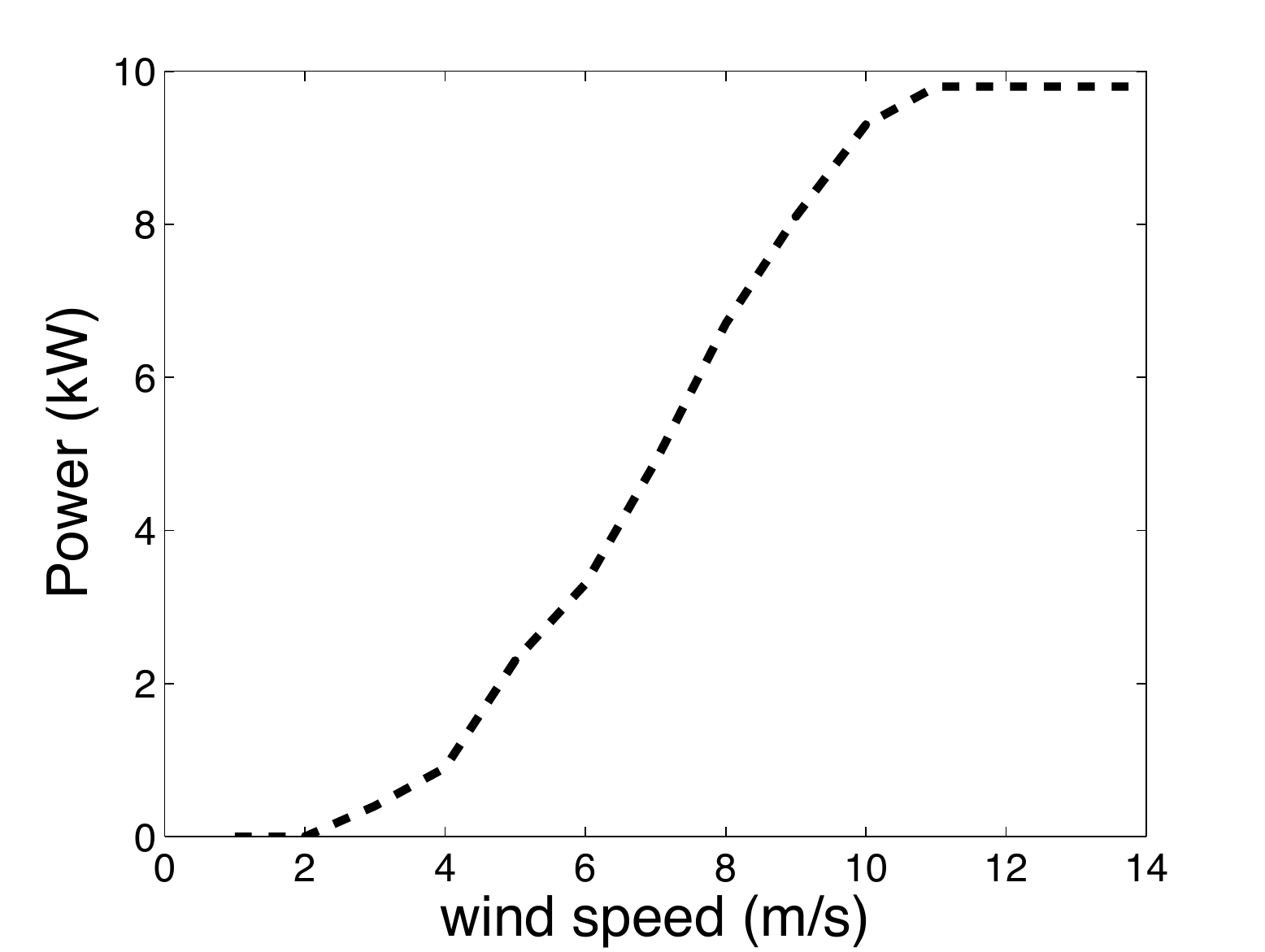}
\caption{Power curve of the 10 kW Aircon wind turbine}\label{pc1}
\end{figure}

In Figure \ref{fig1} we show the real database and its probability density function that follows a Weibull distribution with the following form:

$$ f(v; \lambda, k)=\frac{k}{\lambda} \left( \frac{v}{\lambda} \right)^{k-1} e^{-(\frac{v}{\lambda})^k} $$

where $k$ is the shape parameter and $\lambda$ is the scale parameter. The two parameters have been estimated by a maximum likelihood procedure and, for our database are $\lambda=2.18$ and $k=1.58$. We also estimated the two parameters for the transposed database, we found that the shape parameter remains the same while the scale parameter is $\lambda=10.91$.
The semi-Markov model we use in our analysis is based on a discrete state space which means that uses only discretized wind speed. For this reason we select 8 discretized wind speed states (see Table \ref{st}) chosen to cover all the wind speed distribution. Table \ref{st} shows the wind speed states with their related wind speed ranges.
\begin{table}
\begin{center}
\begin{tabular}{|c|*{2}{c|}|}
     \hline
Sate & Wind speed range $m/s$  \\ \hline
1 & 0 to 3  \\ \hline
2 & 3 - 4  \\ \hline
3 & 4 - 5  \\ \hline
4 & 5 - 6 \\ \hline
5 & 6 - 7  \\ \hline
6 & 7 - 8 \\ \hline
7 & 8 - 9  \\ \hline
8 & $>$9  \\ \hline
\end{tabular} 
\caption{Wind speed discretization}
\label{st} 
\end{center}
\end{table}
The transformation between wind speed and produced energy is made through a commercial 10 kW Aircon HAWT wind turbine. The power curve of this turbine is plotted on Figure \ref{pc1}. The power curve of a specific wind turbine represents how it produces energy as a function of the wind speed values. 

\section{The indexed semi-Markov chain model}
The general formulation of the ISMC model has been developed in \cite{dam1,dam2,dam3,wind2} here we only discussed it informally.

Semi-Markov processes have similar idea as those that generate Markov processes. The processes are both described by a set of finite states $v_n$ whose transitions are ruled by a transition probability matrix. The semi-Markov process differs from the Markov process because the transition times $T_n$ are generated according to random variables. Indeed, the time between transitions $T_{n+1}-T_n$ is random and may be modeled by means of any type of distribution functions.
In studies concerning wind speed modeling the states $v_n$ indicates discretized wind speed at the nth transition and $T_n$ the time in which the nth change of wind speed occurs. 

In (D'Amico et al, 2013), different semi-Markov models were applied to wind speed modeling and it was shown that the semi-Markov models over perform the Markov models and therefore they should be preferred in the modeling of wind speed. 

In order to better represent the statistical characteristics of wind speed, in a recent article, the idea of an ISMC model was advanced in the field of wind speed, see \cite{wind2}. 
The novelty, with respect to the semi-Markov case, consists in the introduction of a third random variable defined as follow:
\begin{equation}
U_{n}^{m}= \sum_{k=0}^{m} v_{n-1-k} \cdot \frac{T_{n-k}-T_{n-1-k}}{T_{n}-T_{n-1-m}}. 
\end{equation}
This variable can be interpreted  as a moving average of order $m+1$ executed on the series of the past wind speed values $(v_{n-1-k})$ with weights given by the fractions of sojourn times in that wind speed $(T_{n-k}-T_{n-1-k})$ with respect to the interval time on which the average is executed $(T_n-T_{n-1-m})$.
The ISMC model considers that the probability of changes in wind speed do depends also on this new variable. In a simple semi-Markov model it depends only on the present wind speed and on the transition time. The ISMC model can be considered as an $m$ order semi-Markov chain where the dependence is given only by the averaged last $m$ states.
Also the process $U^m$ has been discretized, Table \ref{Um} shows the states of the process and their values.
\begin{table}
\begin{center}
\begin{tabular}{|c|*{2}{c|}|}
     \hline
Sate & $U^m$ range $m/s$  \\ \hline
1 & 0 to 2.1  \\ \hline
2 & 2.1 - 2.6  \\ \hline
3 & 2.6 - 3.4  \\ \hline
4 & 3.4 - 6 \\ \hline
5 & $>$6  \\ \hline
\end{tabular} 
\caption{$U^m$ processes discretization}
\label{Um} 
\end{center}
\end{table}

The parameter $m$ must be optimized as a function of the specific database. The optimization is made by finding the value of $m$ that realize the minimum of the root mean square error (RMSE) between the autocorrelation functions (ACF) of real and simulated data, see \cite{wind2}. In our analysis $m=7$.
The reasons to introduce this index of memory are found in the presence of a strong autocorrelation that characterize the wind speed process.
In the same work we have shown that if a too small memory is used, the autocorrelation is already persistent but decreases faster than real data. With a longer memory the autocorrelation remain high for a very long period and also its value is very close to that of real data. If $m$ is increased further the autocorrelation drops again to small values. This behavior suggests the existence of an optimal memory $m$. In our opinion one can justify this behavior by saying that short memories are not enough to identify in which status (low, medium low, medium, medium high, high, see Table 2) is the index $U^m$, too long memories mix together different status and then much of the information is lost in the average.  

The one step transition probability matrix can be evaluated by considering the counting transition between the three random variables considered before. Then, the probability $p_{i,j}(t,u)$ represents the transition probability from the actual wind speed state $i$, to the wind speed state $j$, given that the sojourn time spent in the state $i$ is equal to $t$ and the value of the process $U^m$ is $u$. These probabilities can be computed as:

\begin{equation}
p_{i,j}(t,u)= \frac{ n_{i,j} (t,u) }{\sum\limits_{j} n_{i,j}(t,u)},
\end{equation}
\label{pri}

\noindent where $n_{ij}(t,u)$ is the total number of transitions observed in the database from state $i$ to state $j$ in next period having a sojourn time spent in the wind speed $i$ equal to $t$ and the value of the index process  equal to $u$.

\indent The ISMC model revealed to be particularly efficient in reproducing together the probability density function of wind speed and the autocorrelation function, see \cite{wind2}.

\section{Wind speed forecasting}
In this section the ISMC model is used to forecast future wind speed states by using a one step ahead forecasting procedure,  for different time horizons and for various time scales. Particularly, we tested our model using the previously described databases with a sampling frequency of 10 minutes, 30 minutes, 1 hour and 2 hours. 

For each one of the sampling frequencies, the database is divided into two subsets: the first part is used to find the transition probability matrix (as described in the previous section), we will call this part the setting period; the second part is used to compare the model forecasting with real data (called testing period).
As a first attempt to verify the model performance, we used two years of data as setting period and one year as testing.
Once the transition matrix is set, the forecasted states are computed as follows:
\begin{equation}\label{for}
v^f=\sum_{j=1}^{k} j \, p_{i,j}(t,u),
\end{equation}
where $k$ is the number of states in which wind speed is discretized and $p_{i,j}(t,u)$ is the transition probability matrix. The formula represents the expected value of the next transition given that the present wind speed value is $i$, the sojourn time spent in the state $i$ is equal to $t$ and the value of the index process $U^{m}$ is $u$. 

\begin{figure}
\centering
\includegraphics[height=9cm]{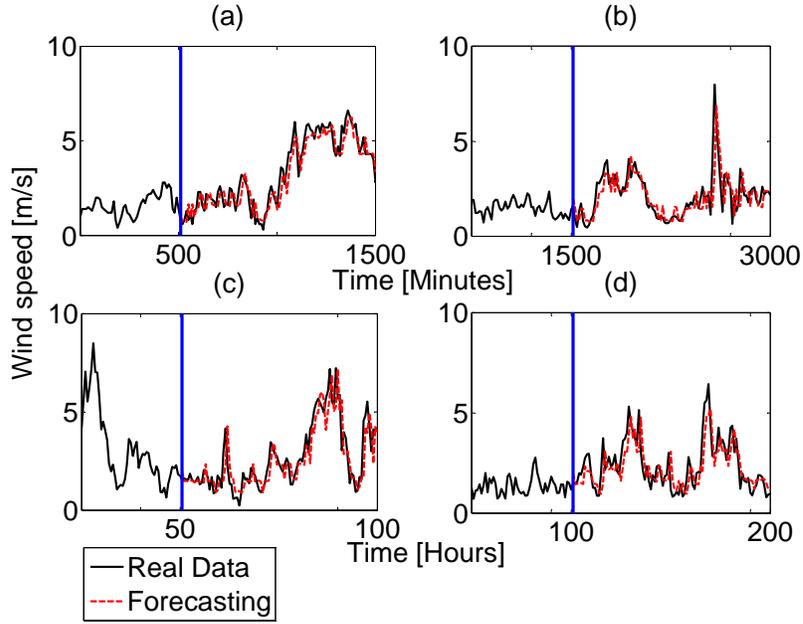}
\caption{Wind speed forecasting one step ahead for 100 time horizon. (a) 10 minutes database, (b) 30 minutes database, (c) 1 hour database, (d) 2 hours database.}\label{figone}
\end{figure}

In Figure \ref{figone} we show the results obtained using our model for the four different time scales. In the figure the black continuous line represents real data while the dashed red line is the predicted series. In this figure the predicted series are long 100 time horizon (specific time depending on the sampling frequency). 

Already from this figure, it is possible to note that the goodness of the prediction does not fall down at the increasing of the length of the forecasted series. This is an important result comparing our model with others used in literature for the same purpose. Once the model is set, for a specific site, the forecasting can be made for a long time.
To better verify this point, in the Table \ref{tab1}  we show quantitative results of our forecasting model for all the considered time horizons. This table shows the mean values and the standard deviations of the RMSE and the MAE made on 50 different forecasted series for a certain time scale. This table shows that the goodness of prediction remains almost constant even varying time scales and time horizons.

\begin{footnotesize}

\begin{table}
\begin{center}

\begin{tabular}{|l|c|c|c|c|}\hline
\diaghead{\theadfont Diag ColumnmnHead II}%
{Indicator}{Time\\Horizon}&
\thead{50}&\thead{100}&\thead{500}& \thead{1000}\\
\hline
RMSE (10 minutes) & 0.44 $\pm $ 0.02 & 0.44 $\pm $ 0.02 & 0.48 $\pm $ 0.02 & 0.52 $\pm $ 0.02 \\
\hline
MAE (10 minutes) & 0.38 $\pm $ 0.01 & 0.37 $\pm $ 0.007 & 0.38 $\pm $ 0.004 & 0.39 $\pm $ 0.003 \\
\hline
RMSE (30 minutes) & 0.48 $\pm $ 0.01  & 0.50 $\pm $ 0.01  & 0.56 $\pm $ 0.01 & 0.62 $\pm $ 0.01 \\
\hline
MAE (30 minutes) & 0.37 $\pm $ 0.01  & 0.37 $\pm $ 0.009  & 0.38 $\pm $ 0.004 & 0.39 $\pm $ 0.003 \\
\hline
RMSE (1 hour) & 0.54 $\pm $ 0.01 & 0.54 $\pm $ 0.01 & 0.61 $\pm $ 0.01 & 0.64 $\pm $ 0.01 \\
\hline
MAE (1 hour) & 0.41 $\pm $ 0.01 & 0.42 $\pm $ 0.01 & 0.43 $\pm $ 0.005 & 0.44 $\pm $ 0.004 \\
\hline
RMSE (2 hours) & 0.56 $\pm $ 0.01 & 0.59 $\pm $ 0.01 & 0.65 $\pm $ 0.01 & 0.69 $\pm $ 0.01 \\
\hline
MAE (2 hours) & 0.48 $\pm $ 0.01 & 0.47 $\pm $ 0.01 & 0.48 $\pm $ 0.007 & 0.48 $\pm $ 0.006 \\
\hline
\end{tabular}
\end{center}
\caption{RMSE and MAE between real wind speed and forecasted series for different time horizon and time scale. The values are expressed in m/s.}
\label{tab1}
\end{table}
\end{footnotesize}

We compare our model with a simple persistence model. This simple method is often used for its simplicity and for its efficiency for very short-term predictions. It assumes that the wind speed at time $t+ \Delta t$ is equal to the wind speed at time $t$. Commonly this method is used to compare the behavior of new forecasting models \cite{sau}. Overall our model has a higher efficiency in the forecast for all the time scales and time horizons. The persistence model does not change its goodness of forecasting at varying of the time horizon. Then we compare our results with the persistence model at different time scales. In Table \ref{tab2} there are the values of the RMSE and MAE for the different time scales and time horizons. As it is possible to note the persistence model has less precision on the forecasting of the wind speed with respect to our model. 

\begin{footnotesize}
\begin{table}
\begin{center}
\begin{tabular}{|l|c|c|}\hline
\diaghead{\theadfont Diag ColumnmnHead II}%
{Time Scale}{Indicator}&
\thead{RMSE}&\thead{MAE}\\
\hline
10 minutes & 0.59 $\pm$ 0.05 & 0.51 $\pm$ 0.01  \\
\hline
30 minutes & 0.63 $\pm$ 0.08 & 0.54 $\pm$ 0.01 \\
\hline
1 hour & 0.73 $\pm$ 0.09 & 0.57 $\pm$ 0.02 \\
\hline
2 hours & 0.85 $\pm$ 0.11 & 0.63 $\pm$ 0.02  \\
\hline

\end{tabular}
\end{center}
\caption{RMSE and MAE between real wind speed and forecasted series for different time horizon and time scale using the persistence model. The values are expressed in m/s.}
\label{tab2}
\end{table}
\end{footnotesize}

\section{Economic application}
An important aspect of wind speed model is that they can be used to forecast the economic validity of an investment in wind farm. We here show how the ISMC model can be used with this aim. We choose to evaluate the Internal Rate of Return, the Duration and the Convexity of the investment on a wind farm in the chosen specific site. We conduced two separate analysis: in the first one we consider an investment for a period of 15 years, which corresponds to the period in which the government incentives on the energy price are still present. In the second case we consider an investment without the government incentives and in this case, being the price of the energy lower than in the previous case, the period of the investment is greater and equal to 30 years.  For the analysis we use the database of wind speed described before. Given that it is composed of only 5 years of wind speed we apply a bootstrap procedure in order to have the sufficient number of data for the analysis. The database has been adapted to the blade height by using the mother described in equation \ref{height}.
To compare real data with the ISMC model we generate, through Monte Carlo simulation, 1000 possible scenarios of wind speed time series of length 15 years for the firs case and 30 years for the second case, the sampling period is 10 minutes.

\subsection{Energy production}
According to the power curve described in Figure \ref{pc1} wind speed has been transformed into power, and then, in produced energy at a specific time window.
In Figure \ref{enr1} it is shown the comparison between the distribution of the produced energy in a 10 minutes interval for real and synthetic data. Being the transformation from wind speed into energy not linear, this comparison shows that our model is able to reproduce correctly also a real case of energy production. In fact the values obtained in the two cases are almost identical. The particular form of these distributions is due to the existence of a cut-in wind speed, under which no energy is produced, and a part of the power curve that remain with constant value, from 10 $m/s$ until 32 $m/s$, in which the wind turbine produce energy at its rated power. This explain the existence of the two modes.

\begin{footnotesize}
\begin{figure}
\centering
\includegraphics[height=6cm]{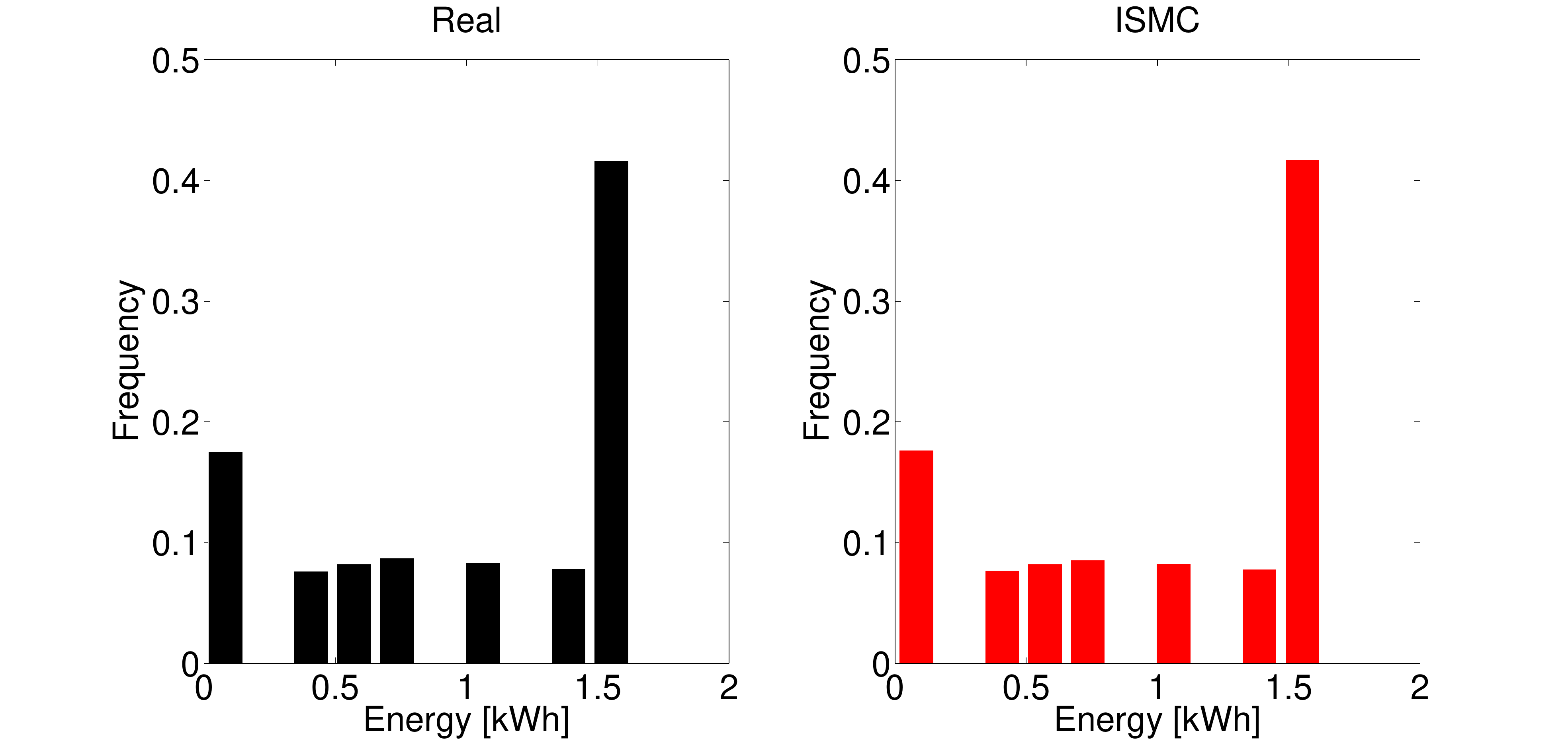}
\caption{Histogram of the electrical energy produced in 10 minutes. Comparison between real and simulated data.}\label{enr1}
\end{figure}

\begin{figure}
\centering
\includegraphics[height=6cm]{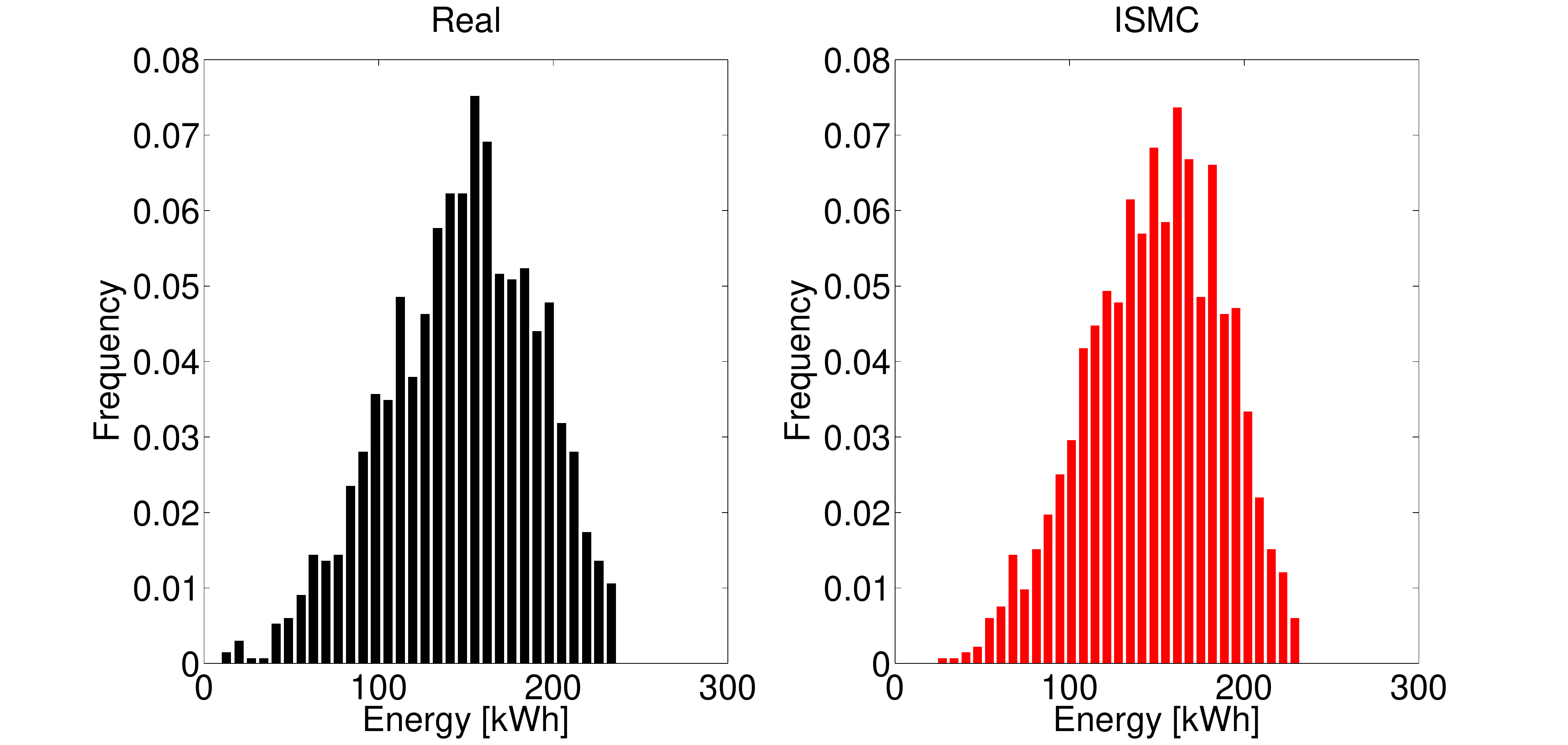}
\caption{Histogram of the electrical energy produced in one year. Comparison between real and simulated data.}\label{enr2}
\end{figure}
\end{footnotesize}

In Figure \ref{enr2} we show the distribution of energy produced in 1 year. The comparison is made again between real and synthetic data. The mean value and the standard deviation of the real annual energy produced are $147.4 \pm 42.9 \; kWh$ instead for the simulated case we have $148.8 \pm 38.5 \; kWh$. The cut part of the right side of each histogram is due to the presence of a cut-off wind speed, that is the value of the wind speed at which the wind turbine is stopped for structural reason. For wind speed over this value there is no production of electrical energy.

\subsection{Financial indicators and results}
We will compare some financial indicators computed on the real time series, generated through the bootstrapping procedure explained before, and the simulated date generated through the ISMC model by means of Monte Carlo simulation. Just to remind, we are going to consider two different cases. In the first case we compute the cash flow obtained by the selling of energy, produced by the wind turbine, taking into account the government incentives on the selling price. In Italy for a wind farm with a rated power under 1 $MW$ and until 15 years from the setup of the plant, the price of the energy is fixed, at the time of writing, to 0.291 \euro$/kWh$. In this case we consider a cash flow of 15 years. The second case takes into account the selling of energy at the market price, without the government incentives. In this second case we fix the energy selling price at 0.027 \euro$/kWh$ and, being it smaller than the price with the incentives, the considered period of the cash flow is extended to 30 years.

For each case we generate 1000 possible scenarios and we evaluate the financial indicators for both real and synthetic data. The trajectories are transformed into power through the power curve of the wind turbine chosen, then in energy by considering the time between two successive steps and after in euro by multiplying for the specific price considered. The annual cash flow are built and at time $0$ is considered the coast of the initial investment. Commonly for wind turbines with a rated power under 100 $kW$ it is considered a price of 3000 \euro  for each $kW$. This price takes into account the main cost of the turbine, the setup cost and also the annual maintenance. Being our wind turbine a 10 $kW$ of rated power, we can easily choose as initial investment 30000 \euro. At last the financial indicators for all the scenarios are evaluated to compare the results obtained by real data and the synthetic one. The indicators chosen are the IRR (internal rate of return), the Duration and the Convexity. The IRR is the annualized effective compounded return rate that makes the net present value of all cash flows equal to zero, then it can be evaluated as follow:
$$
\sum_{s=1}^{n} CF_s \cdot (1+r)^{-s}=0
$$

Where $CF$ is the cash flow of the year $s$, $r$ is the internal rate of return and $n$ is the number of years of the investment. 
The second indicator (Duration) represents the length of the mean life of the cash flow of an investment weighted with their actualized values and can be formalized as follow:

$$
D=\frac{\sum_{s=1}^{n} s \cdot CF_s \cdot (1+r)^{-s} }{ \sum_{s=1}^{n} CF_s \cdot (1+r)^{-s}}
$$

The last financial indicator is the Convexity that is a measure of the sensitivity of the duration of an investment to changes in interest rates. This indicator can be evaluated with the following formula:

$$
C=\frac{1}{(1+r)^{2}} \cdot \frac{\sum_{s=1}^{n} s \cdot (s+1) \cdot CF_s \cdot (1+r)^{-s} }{ \sum_{s=1}^{n} CF_s \cdot (1+r)^{-s}}
$$

The notation inside the two previous formulas of the Duration and Convexity are the same of the IRR with the difference of the interest rate $r$ that in these two last cases we consider equal to 3\%.

Figures \ref{hist1}, \ref{hist2}, \ref{hist3} show the histograms of the financial indicators evaluated for all the 1000 scenarios, for real and simulated data, and for the two cases with and without the government incentives. As it is possible to note, for each pair of indicators and for each specific case of the period of investment, the mean values of the indicators are almost the same for real and simulated data.

\begin{footnotesize}
\begin{figure}
\centering
\subfloat[]{\label{main:v}\includegraphics[height=6cm]{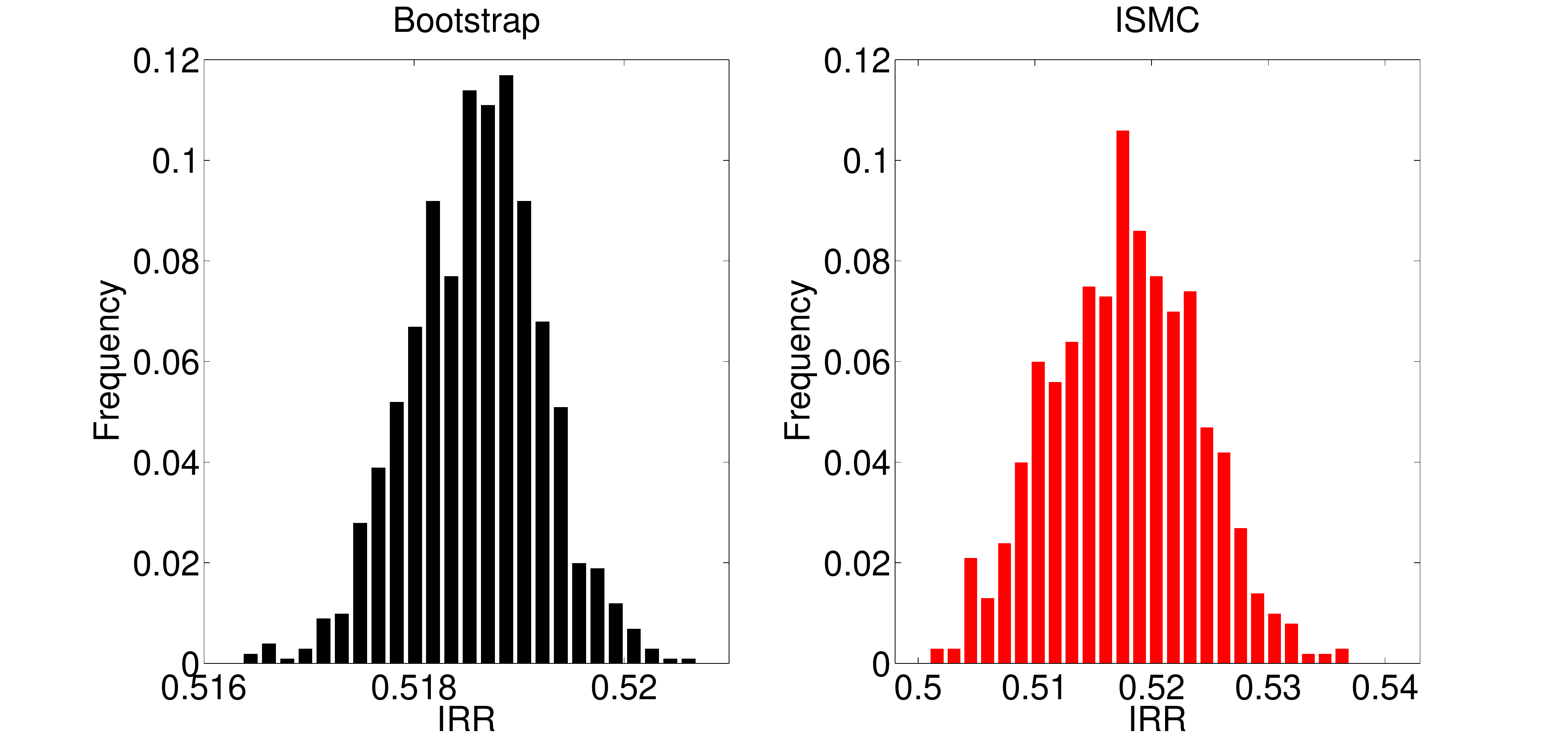}}

\subfloat[]{\label{main:b}\includegraphics[height=6cm]{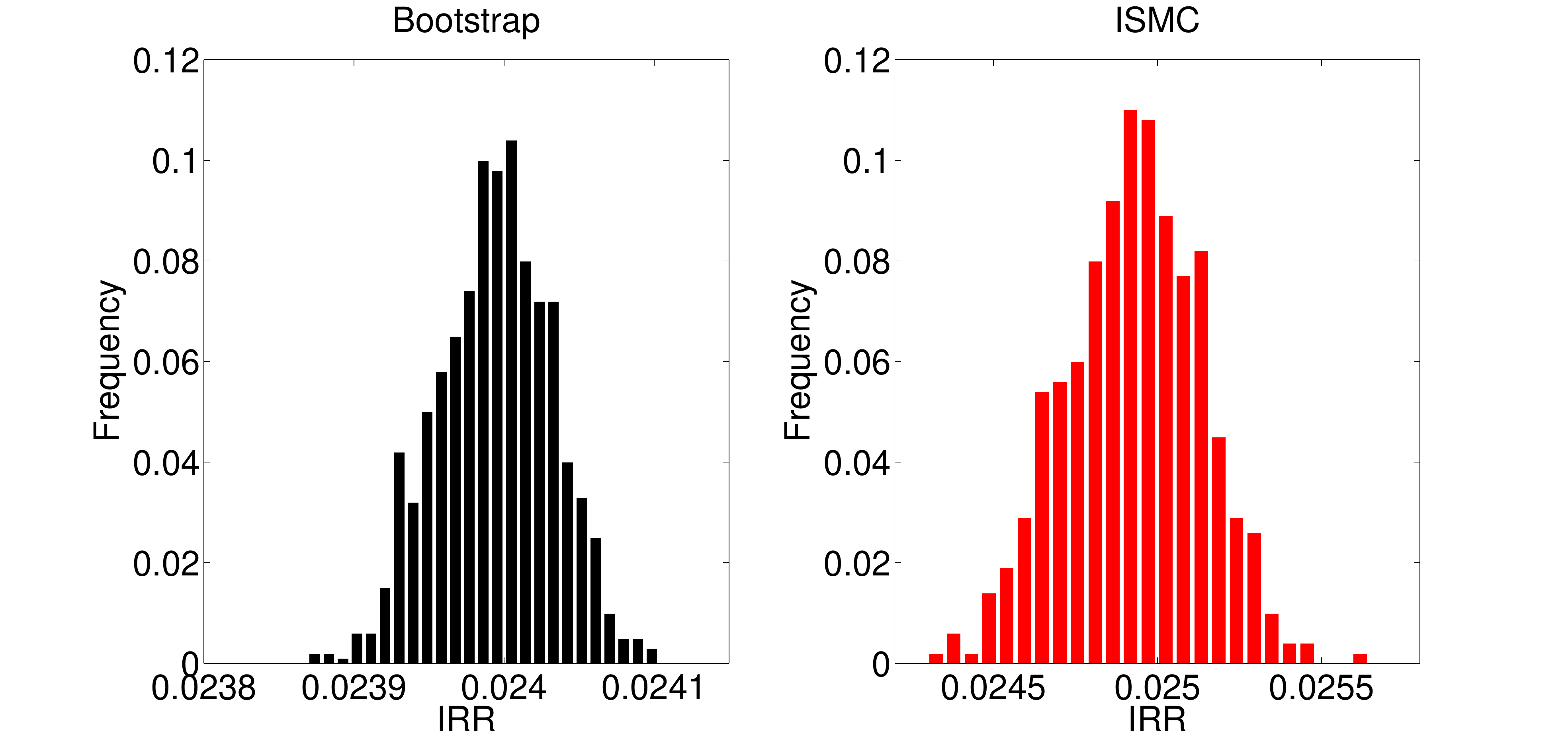}}
\caption{Histogram of the Internal Rate of Return for real and simulated data. (a) Case with government incentives and cash flow of 15 years. The mean value and standard deviation for the ISMC and bootstrap series are respectively 0.5256$\pm$0.0063 and 0.5186$\pm$.0062. (b) Case without government incentives and cash flow of 30 years. The mean value and standard deviation for the ISMC and bootstrap series are respectively 0.0249$\pm$.0021 and 0.0240$\pm$0.0003.}
\label{hist1}
\end{figure}

\begin{figure}
\centering
\subfloat[]{\label{main:n}\includegraphics[height=6cm]{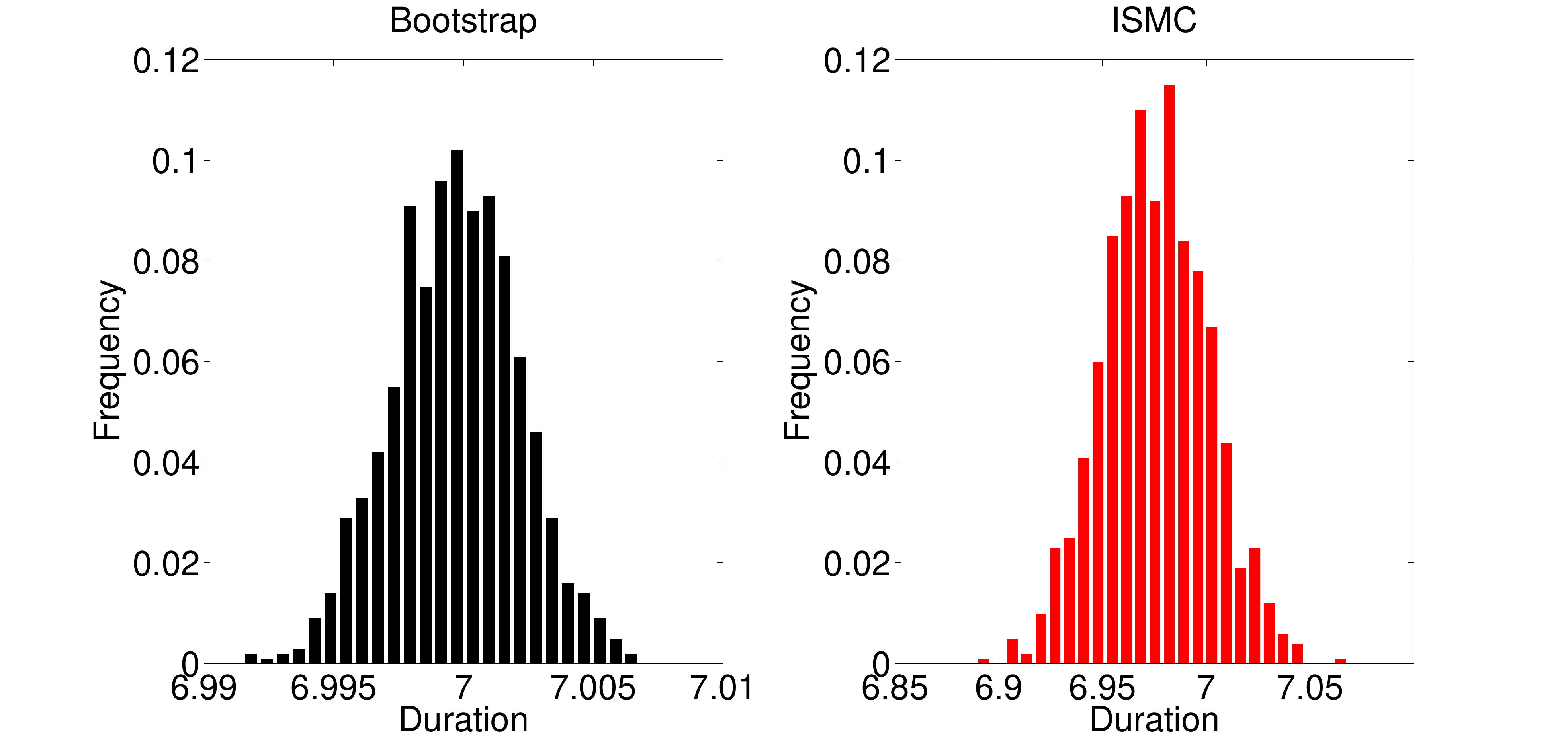}}

\subfloat[]{\label{main:m}\includegraphics[height=6cm]{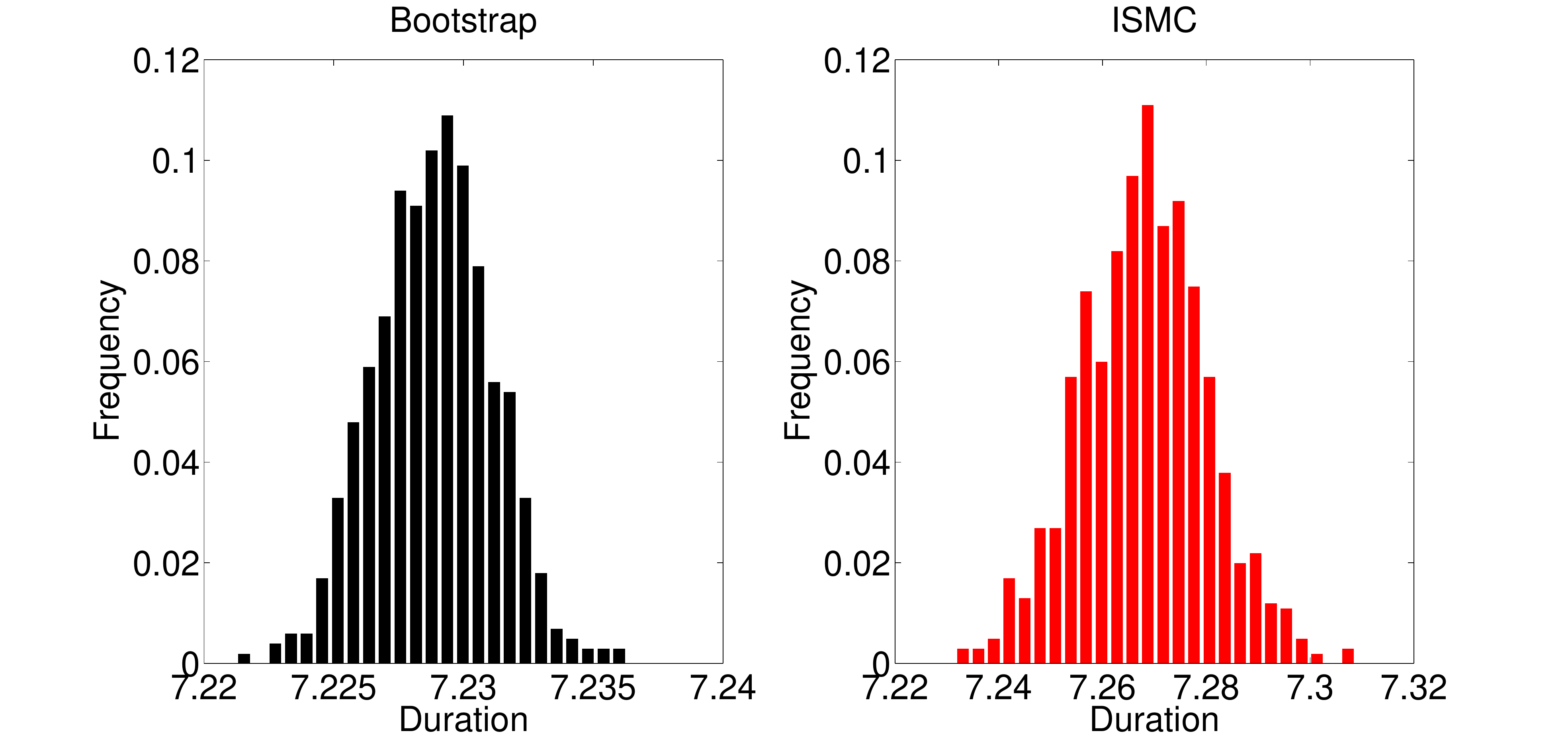}}
\caption{Histogram of the Duration for real and simulated data. (a) Case with government incentives and cash flow of 15 years. The mean value and standard deviation for the ISMC and bootstrap series are respectively 7.0134$\pm$.0253 and 6.9997$\pm$0.0024. (b) Case without government incentives and cash flow of 30 years. The mean value and standard deviation for the ISMC and bootstrap series are respectively 7.2680$\pm$.0122 and 7.2288$\pm$0.0023.}
\label{hist2}
\end{figure}

\begin{figure}
\centering
\subfloat[]{\label{main:a}\includegraphics[height=6cm]{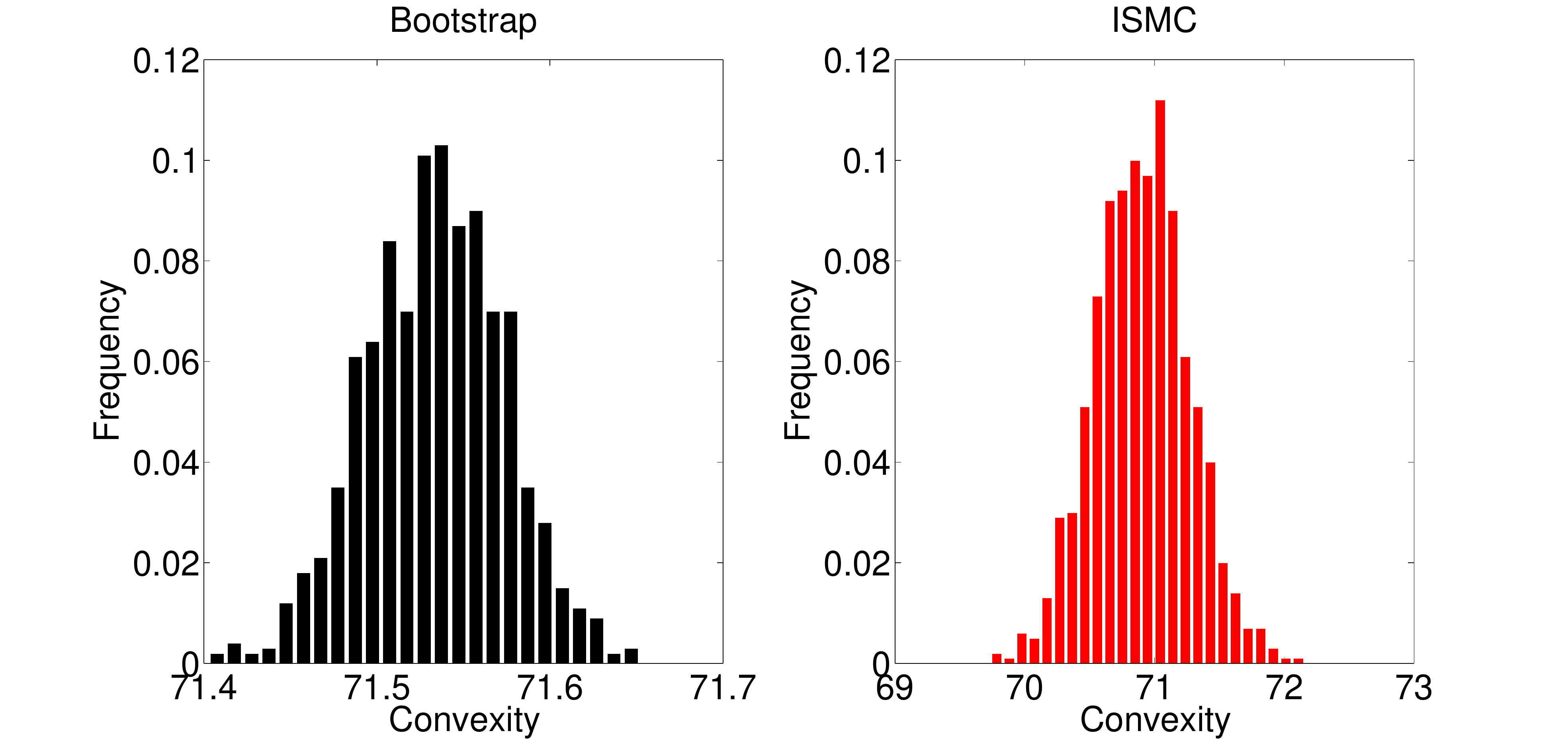}}

\subfloat[]{\label{main:c}\includegraphics[height=6cm]{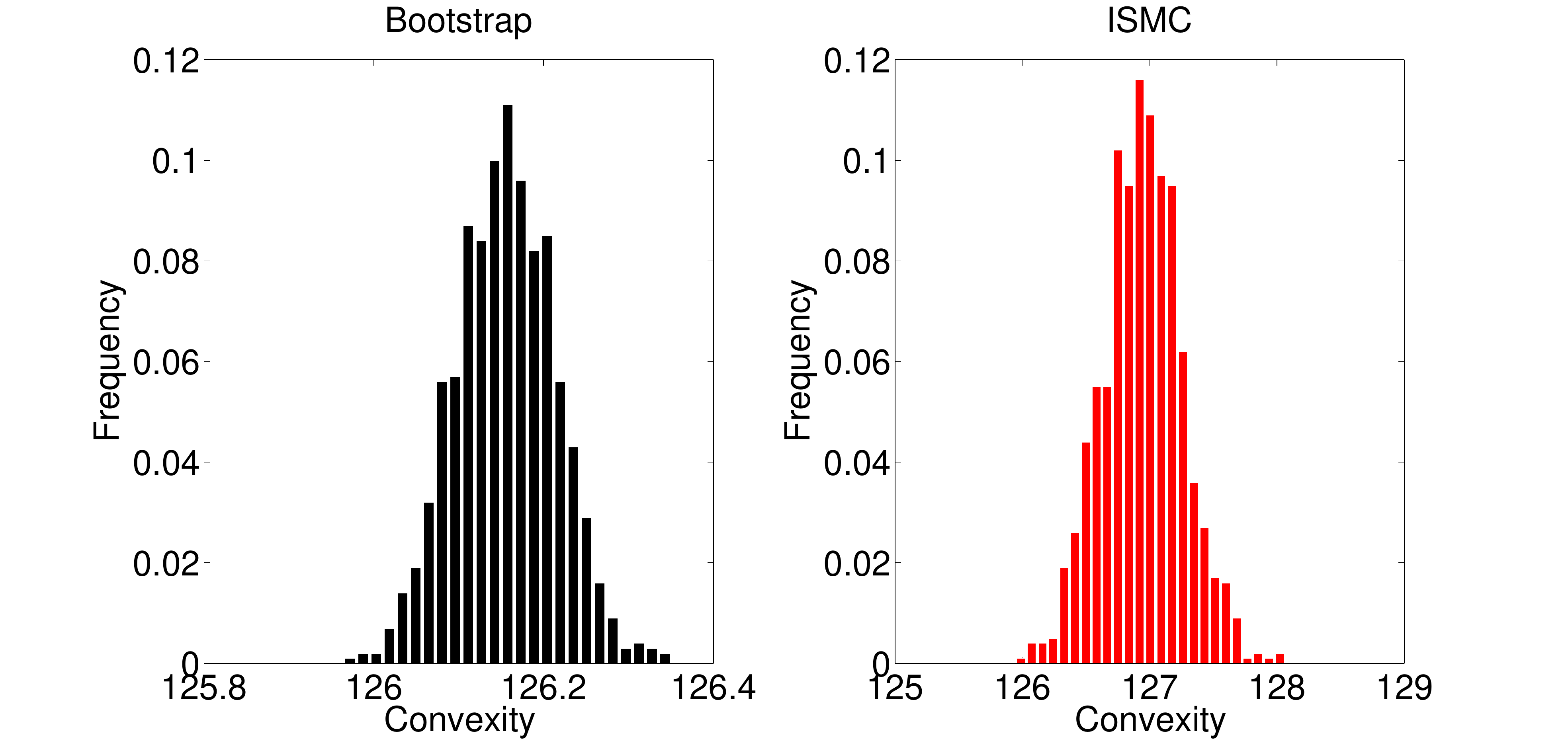}}
\caption{Histogram of the Convexity for real and simulated data. (a) Case with government incentives and cash flow of 15 years. The mean value and standard deviation for the ISMC and bootstrap series are respectively 71.7516$\pm$0.3638 and 71.5337$\pm$0.0401. (b) Case without government incentives and cash flow of 30 years. The mean value and standard deviation for the ISMC and bootstrap series are respectively 126.9422$\pm$0.3119 and 126.1550$\pm$0.0590.}
\label{hist3}
\end{figure}
\end{footnotesize}

\section{Discussion and conclusion}
In previous works we presented new stochastic models, all based on a semi-Markov approach, to generate synthetic time series of wind speed. 
We showed that all the models perform better than corresponding Markov chain based models in reproducing statistical features of wind speed. Using these results, here, we tried to apply the model which we recognized to be the best among those, namely the indexed semi-Markov chain (ISMC) model, to forecast future wind speed in a specific site. 
The ISMC model is a nonparametric model and because of this it does not need any assumption on the distribution of wind speed and on wind speed variations.

In previous papers we showed that the ISMC model is able to reproduce correctly, and at the same time, both the probability distribution function of wind speed and the autocorrelation function. 

The results presented in this paper show that the model can be efficiently used to forecast wind speed at different time horizons. The forecast performance is almost independent from the time horizon used to forecast; the model can be used without degradation during the considered horizon time, at different time scales (we showed this for time scales ranging from 10 minutes to 2 hours).
The number of data needed to reach a good forecast performance do depend on the time scale used for forecasting; the model always works better than a simple persistence model.

The financial analysis shows that our model can be used to perform the evaluation of a specific site for the investment of a wind farm with a limited number of sampling data. 

All these characteristics suggest that the advanced ISMC model may be used both for modeling wind speed data and for wind speed prediction. Therefore, it may be utilized as input data for any wind energy system.

\end{document}